# *Reasoning within and between collective action problems*

Ofer Tchernichovski[1], Seth Frey[2], Dalton C. Conley[3] & Nori Jacoby[4]

1. otcherni@hunter.cuny.edu, Department of Psychology, Hunter College, CUNY New York, NY 10065 USA. 2. sethfrey@ucdavis.edu Department of Communication, University of California, Davis Davis, CA 53706 USA. 3.Department of Sociology, Princeton University, Princeton NJ. 08544 USA 4. kj338@cornell.edu, Department of Psychology, Cornell University Ithaca, NY 14853 USA

**Abstract**

Understanding cooperation in social systems is challenging because the ever-changing rules that govern societies interact with individual actions, resulting in intricate collective outcomes. In virtual-world experiments, we allowed people to make changes in the systems that they are making decisions within and investigated how they weigh the influence of different rules in decision-making. When choosing between worlds differing in more than one rule, a naïve heuristics model predicted participants' decisions as well, and in some cases better, than game earnings (utility) or by the subjective quality of single rules. In contrast, when a subset of engaged participants made instantaneous ("within-world") decisions, their behavior aligned very closely with objective utility and not with the heuristics model. Findings suggest that, whereas choices between rules may deviate from rational benchmarks, the frequency of real time cooperation decisions to provide feedback can be a reliable indicator of the objective utility of these rules.

**Keywords:** collective action; feedback systems; collective intelligence; social dilemmas; self-governance

## Introduction

Modern human society is founded on massive social networks and institutions that require cooperation in complex environments (Ostrom, 2015; Tchernichovski et al., 2023). Cooperation is often challenged by constraints interacting over multiple timescales, particularly as agents participate in changing the "rules of the game" governing those interactions (Frey & Atkisson, 2020). Do people's intuitive "folk" theories of effective collective behavior accurately predict cooperation outcomes? How does that accuracy change with the complexity of the community's regulatory environment? Collective intelligence scholars are increasingly interested in treating collectives as if they were cognitive agents (Goldstone & Gureckis, 2009; Couzin, 2009; Malone & Bernstein, 2025), just as cognitive scientists are taking an increasing interest in social institutions (Misyak et al., 2016; Hawkins et al., 2019; Wu et al., 2024). With this interest, many familiar frameworks find new grounding.

In an analogue to Marr's implementation and algorithmic levels of analysis (Frey & Atkisson, 2020; Marr et al., 2010), political (and cognitive) scientist Elinor Ostrom introduced a distinction within governance institutions between the operational and collective-choice levels of analysis (Kiser & Ostrom, 1982; Ostrom, 2019). Within this framework the <u>operational level</u> articulates how to implement the operations necessary to the functioning of the system, while the <u>collective-choice level</u> arranges those operations such that individuals' behaviors are appropriately integrated to make the behavior of the collective coherent. In our experiments, the <u>collective-choice</u> level is where participants choose between virtual world games with different rules, governing cooperation. Instantaneous choices made while interacting with the virtual world (within one of those rule sets) would be at the <u>operational level</u>.

Enabling participants to iteratively influence the rules governing their own cooperation motivates several relevant questions for collective behavior, collective action, and collective intelligence. How accurate are people's forecasts of the collective's behavior under a given set of rules? How consistent are people in supporting cooperation at each level of choice? And how do forecasts change in accuracy with changes in rule complexity? To test these questions, we performed large scale online experiments in virtual worlds. In each world, participants played cooperative games based on combinations of binary rules, which we implemented based on the Ostrom theoretical framework for self-governance (Ostrom, 2009). They alternated between making collective-choice level decisions *between worlds* and operational level decisions *within worlds*. In our first experiments, the rule system was "simple" in the sense that rules could change independently of each other, and all decisions were at the collective choice level of choosing between worlds after some experience playing within different sets of rules. In our second series of experiments, we allowed participants not only to choose whether to opt into a set of rules, but also to decide how much to contribute (at the operational level) once they opted in. With this series of experiments, we were able to give participants experience working together under a wide variety of institutional regimes, providing them the opportunity to participate in which regimes they acted under.

## Decision-making models

To interpret participants' decisions, we developed simple decision-making models and then evaluated how well each model can explain the observed behavior. When people decide between two options, a natural model would be to assume that they make the decision based on the objective or subjective utility value of these options. It is often assumed (Train, 2010; Harrison et al. 2020) that the utility has the form: $U_i = u_i + n_i$, where $n_i$ is a stochastic component distributed with a Gumbel distribution, and $u_i$ is a deterministic component. In this case, the probability $p_{21}$ to choose option 2 over option 1 can be expressed as

1) $p_{21} = \frac{1}{1+e^{-\gamma(u_2-u_1)}}$

where $\gamma$ is a constant that depends on the internal noise. In case that participant uses the true objective utility, this gives us a model to predict $p_{ij}$ (for every two options in the experiment) based on $u$ with a single degree of freedom ($\gamma$) which can be fitted from the data. In addition, under this assumption we can determine the relation between option 3 and 1 ($p_{31}$) if options 3 and 2 ($p_{32}$) and 2 and 1 ($p_{21}$) are known. First, we note that a relation like the following holds true for $p_{21}$, $p_{32}$ and $p_{31}$: $\frac{1}{p_{ij}} - 1 = e^{-\gamma(u_i - u_j)}$

However: $e^{-\gamma(u_3 - u_2)} e^{-\gamma(u_2 - u_1)} = e^{-\gamma(u_3 - u_1)}$

Thus: $(1/p_{31} - 1) = (1/p_{32} - 1)(1/p_{21} - 1)$, which can be reorganized to an elegant formula:

2)  $p_{31} = 1/(2 - 1/p_{32} - 1/p_{21} + 1/(p_{32} p_{21}))$

Assuming that the utility in all games directly corresponds to the objective utility (coin earning rate), the observed frequency of choosing between each game can be compared to the expected frequency according to equation (1). Alternatively, and as observed empirically in some cases, individuals may rely on subjective utility. In this scenario, equation (2) can be used to predict the choice between games 3 and 1 based on the observed frequencies of selecting 3 over 2 and 2 over 1, to evaluate if participants are making rational decisions based on these subjective values. Finally, people may follow a simple heuristic leading to averaging the contribution of rule probability:

3)  $p_{31} = \frac{1}{2}(p_{32} + p_{21})$

which we will refer to as the 'folk's model'.

In experiments where participants are confronted with the option to activate a decision (such as join the club) given a set of rules $w$, calculating the expected frequencies is a bit different. Here, we have two relevant utilities:

$u_{\text{join},w}$ and $u_{\text{opt-out},w}$ If people can experience both options, and are exposed to these utilities, one natural choice would be for them to be guided by these objective gains. In this case we expect them to behave according to equation 1 and the real utilities (e.g. coin earning).

Finally, in situations where the rational decision would be to avoid cooperation entirely, e.g., free riding (Tchernichovski et al., 2023), individuals often cooperate at levels higher than the expected Nash equilibrium, which predicts no cooperation (Mao et al., 2017). In such situations in our experiments, we empirically found that people heuristically follow the utility when they decide to cooperate - namely the utility and cooperation rate are highly correlated. We thus hypothesize that they still follow a relation similar to equation (1) namely the probability to cooperate $p_C$ follows this relation:

4)  $p_C = \frac{1}{1 + e^{-\gamma u_w}}$

where $u_w$ is the utility given the set of rules $w$ in the game.

## Methods

### Participants

We recruited a total of 1101 participants that provided consent in accordance with approved protocol. All participants were recruited online using Prolific and Amazon Mechanical Turk with the following constraints on recruitment: (i) participants must be at least 18 years old, and (ii) have a 95% or higher approval rate on previous tasks. Participants were paid at a US $10/hour rate according to how much of the experiment they completed in addition to a performance bonus based on the number of coins they collected in each game (1 cent per coin, total bonus between $1-$3 depending on performance). The completed experiments took approximately 10-25 minutes.

### *Automated Online Implementation*

The experiments were implemented in PsyNet (https://www.psynet.dev/), a Python package for performing complex online behavioral experiments at large scale (e.g., Harrison et al., 2020). The 3D virtual environment (Unity WebGL app) was integrated with Psynet, which controlled the game logic across experimental conditions, communicating with a backend Python server cluster responsible for organizing the experiment and collecting data.

### *Games design*

All games were coded using Unity 3D (https://unity.com), compiled to WebGL, and were combined with PsyNet, which managed the online implementation via Chrome web browser (see below). All games were implemented asynchronously, such that information (rating scores) posted by previous players accumulated in the world that the next players encountered. All games implemented the native Unity 3D physics, and first-person player design, where the participant sees the game world through the perspective of their avatar's eyes. Before the game, participants watched a one-minute video explaining and demonstrating the game.

We first tested how game rules interact to affect collective choice level cooperation decisions. The naturalistic 3D game provides an experience similar to that of collective action problems in the real world (**Fig. 1A-B**), such as online marketplaces, where people make different gambles (test different arms of a collective multi-armed bandit problem), with the option, but not the obligation, to share their evaluations of each choice with others (Tchernichovski et al., 2019). In the game, participants choose and evaluate service utilities while exploring and gaining profit in the virtual world (these virtual gains were paid to participants as real money at the end of the game). In each turn, participants navigate in virtual islands to search and collect coins (that count toward their monetary reward for participation). Each island had 20 collectable coins, each worth one cent. Participants commuted between islands by riding simulated ferries. Each ferry type corresponds to an arm in the embedded multi-arm-bandit (MAB) problem with a total of nine ferry types (=bandit arms) ranging in speed (Tchernichovski et al., 2023). In each turn, players choose between two arms (ferry types) selected from the overall nine. The bandit arm reward is the speed of the ferry, ranging between 2 and 30 seconds, with faster ferries providing participants more time at the destination to compete in coin collection.

To add a "collective intelligence" dimension to the MAB, we implemented a public ferry rating dashboard, allowing participants to share with each other their own assessments of a given ferry's speed, akin to public product rating on online marketplaces. The slider was costly, such that reporting extreme scores took a few seconds longer than reporting moderate scores (Tchernichovski et al., 2019). In return for the inconvenience and time consumed by rating the ferry, participants enjoyed guidance to selecting faster ferries: a dashboard indicated aggregated rating scores for each ferry (**Fig 1A-B**). To help emphasize the social nature of the setting, and to create time pressure, players' coin collection competed in real time with artificial agents built as Unity AI bots, which behaved in a manner similar to that of a human player: bot avatars were exploring the virtual island, collecting coins, and riding ferries. Participants competed against three AI agents (bots), who collected coins. Choosing a slow ferry places the player at a disadvantage, as bots who ride the fast ferry can start collecting coins before the other participants arrive. The bots selected ferries using the same rating information available to the human participants. Participants were informed that they will play against bots, and that all rating scores presented for ferries were posted by real human participants. Participants could play cooperatively with each other to find the fastest ferries across coin islands. Their cooperation could occur under different combinations of rules. We tested participants' choices for three binary cooperation rules (**Fig 1C**).

With the profit-sharing rule turned on, participants shared 50% of the coins they collected. Profits from the shared coins, plus a dividend, were shared equally between players at the end of each round. The second rule was taxation. When turned on, a 2-coin toll was collected in each ferry ride. In return to the tax, all ferries operated at a faster speed, reducing the average ferry riding time by 10 seconds, from an average of 15 seconds to an average of 5 seconds. The taxation was designed to increase net profits by allowing more time to collect coins, and it also reduced inequality by narrowing the gaps between arrival times of different ferries. The third rule, rating requirements, activated the ferry rating dashboard, requiring participants to provide public ratings of their ferry's speed after each ride. When rating duties are turned off, participants do not rate the ferry and there is no dashboard (**Fig. 1B**). Note that with rating duties turned off, the cohort is no longer solving the MAB problem collectively, and ferry choices are based on personal experience, which we designed to be deficient by shuffling the bandit arms in each world. By designing a rule space with binary rules that can be activated independently of each other, we can estimate how different combinations of rules affect both cooperation and collective intelligence, as the two are not trivially coupled (Tchernichovski et al 2023).

## Results

**Choices between games with non-interacting rules**
We recruited 193 participants and tested their choices in the transition from a game where all rules are turned off, through intermediate stages, up to a game where all rules are turned on (FFF→TTT). Each participant was first assigned to play one game (Game A) for four turns. Then, the participant was assigned to play a second game (Game B) for four turns. Finally, representing a shift from "operational" level to the "collective choice" level, the participant was prompted to choose between Game A and Game B for playing the last four turns (**Fig. 1E**).

The three binary (T/F) rules give eight unique games. **Figure 2A** presents all possible trajectories from FFF to TTT between these games. We decided to test one of these trajectories (orange nodes). We first evaluated the objective utility of different rules by calculating the average earning rate for different games among our participants (**Fig. 2B**). As expected by the game design, turning all rules off resulted in the lowest utility and that turning on resulted in the highest utility. Surprisingly, however, the transition frequency from FFF→TTT was at chance level (0.50, **Fig. 2C**). This outcome could suggest that participants were un-attentive to the rules, or too uncertain about the new environment to risk exploring it. However, transition frequencies *between games* that differed by a single rule show a statistically significant deviation from chance (Chi squared = 11.759, df=5, p=0.0382). This outcome was driven most strongly by the low frequency of FFT→FTT indicating a strong bias against rule 2, paying ferry tax (**Fig. 2C**), despite higher gains (**Fig. 2B**). Consequently, fitting an objective utility to all the decision frequencies gave a negative gamma (model equation (1), **Fig. 2C**), indicating a failure of earning rate to predict decisions.

We therefore reverted to testing if, at the very least, the observed frequencies of choices *between games* that differed by a single rule can predict choices between worlds that differed by more than one rule. That is, can the subjective utility estimates according to single rules, fit the rest of the data when rules are combined (model equation (2), **Fig. 2D**, green triangles). We compared the subjective utility model predictions to those of the heuristic averaging (folk's theory, model equation 3), according to which participants naively average the probabilities (red triangles in **Fig. 2D**). As shown, in all three comparisons, the heuristic averaging model outperformed the subjective utility model. This suggests that heuristic averaging is at least as good (and perhaps a better) predictor of participants' choices between the games.

We wondered if participants would perform better if we tested a trajectory between games that is aligned with the subjective utility we estimated in the previous experiment. We recruited a new cohort of 137 participants, and instead of setting TTT as a target, we set the target as TFT (with taxation off). Instead of FFF as a starting point, we set FTT (with taxation on, and without coin sharing). As shown in **Figure 2E**, the combined rules transition FTT→TFT showed a somewhat, but not significantly better fit to the prediction of subjective utility compared to that of heuristic averaging ($X^2=0.84$, p=0.35).

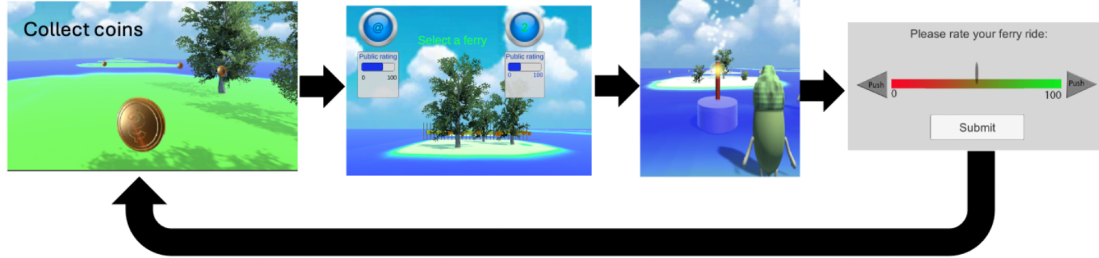
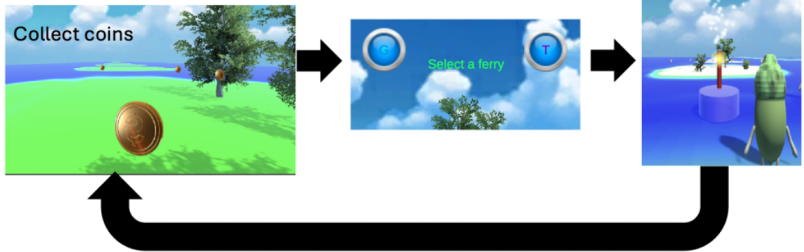
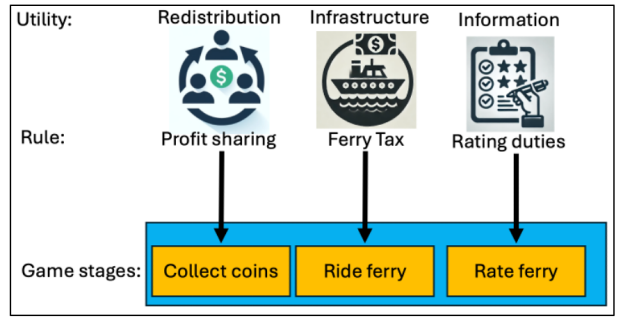
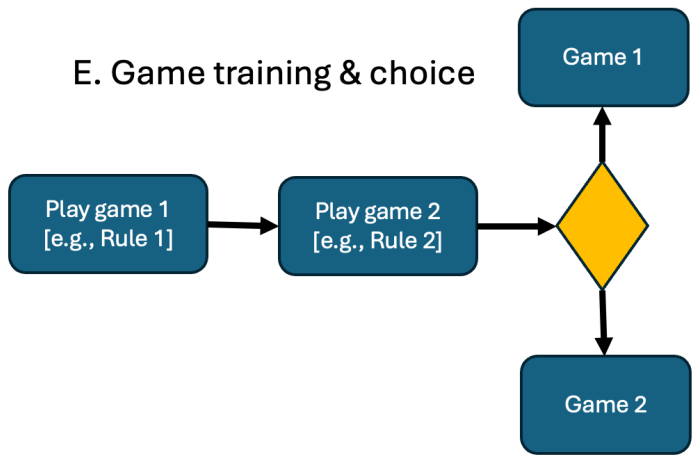

**Figure 1:** *A, game stages: coin collecting in an island (against bots); choice between ferries (with rating dashboard); ferry riding; ferry rating. B, an example of game stages with rating rule turned off. C, binary rules, each correspond to a specific utility and to a specific stage in the game (D). E, experimental timeline. Participants first play one game, then they play a second game, and finally they choose between the two games for the rest of the experiment.*

Each rule represents a distinct utility, affecting a unique stage in the game (**Fig. 1D**):

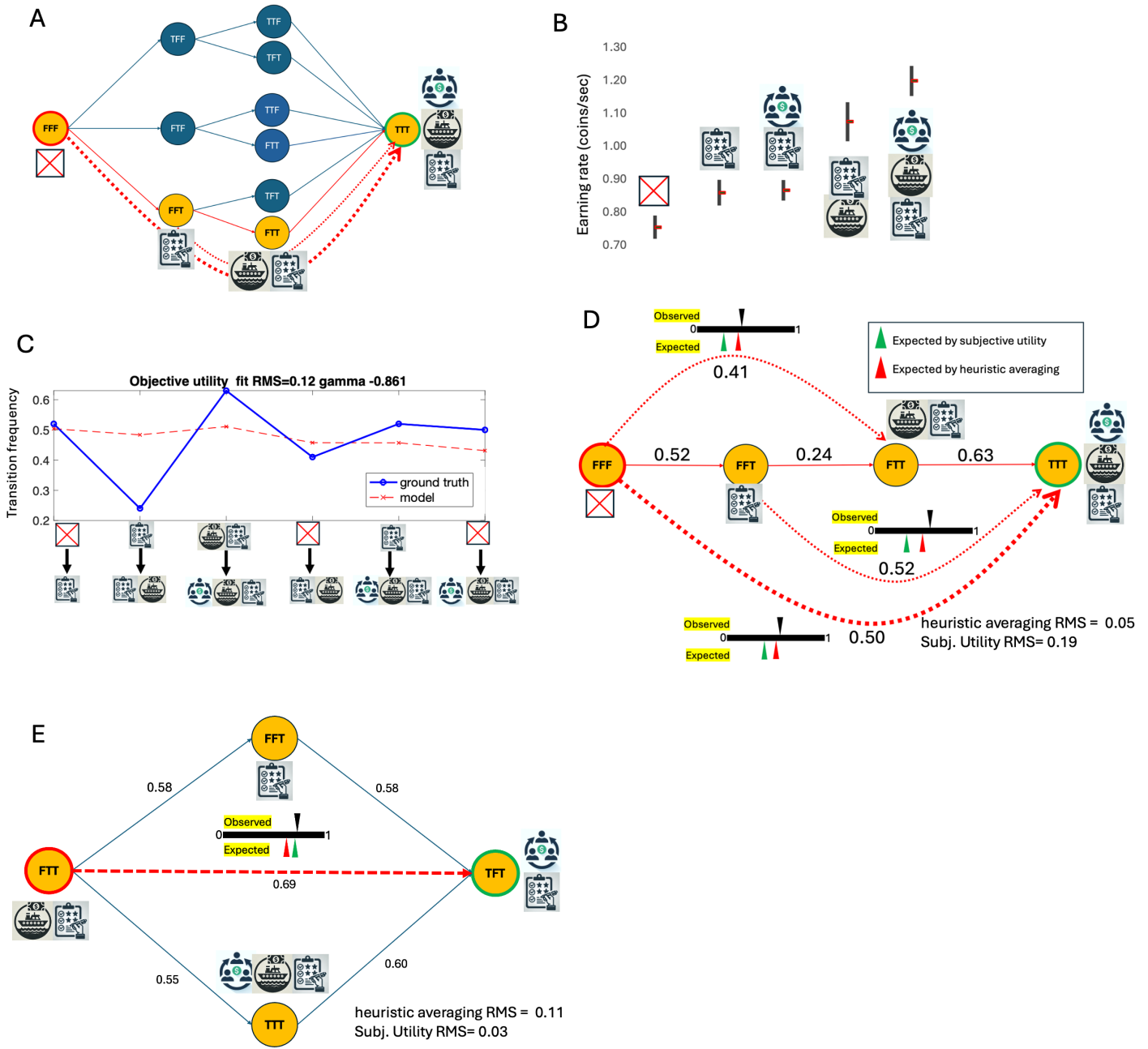

**Figure 2:** *A, All possible transitions from all rules turned off (FFF) to all rules turned on (TTT). Yellow nodes highlight transitions that were tested experimentally. B, Objective utility of rules. C, Fitting of objective utility model to transition frequencies. D, Estimated transition frequencies between games. Expected frequencies for higher order transitions are based on subjective utility (green) and averaging the observed lower order transitions estimates (red). E results from a follow-up experiment guided by subjective utility estimates.*

## Choices between games with nested rules

We next test model predictions in games where rules interact to produce more complex institutional arrangements. We designed a simpler version of the coin collecting game with no competing AI bots. As before, we designed engagement rules that were loosely based on the Ostrom rule typology framework and could vary independently (**Fig. 3A**). Here, however, we designed rules that are contingent upon each other within and across stages of the game (**Fig. 3B**). The position rule assigns participants that collected coins quickly as *Leaders*. Leaders could move faster and collect more coins but were under obligation to rate all ferries. The second rule, payoff rule, incentivized ferry rating by offering a payment for riding and then rating the least known ferry (and in this manner, facilitating the accumulation of ratings for all ferries). The third rule, escape rule, allowed participants to escape from a slow ferry in exchange for paying 3 coins. The fourth rule, performance rule, requires participants to collect at least 80% of the coins in each island before they can move to the next island.

In this design, time pressure came from an overall time limit on participation in the experiment, which incentivized participants to choose faster ferries by leaving them time to visit more islands and collect more coins. We design the game to compare *collective-choice* level (*between games*) to choices made at the *operational level* (*within game*). Between game sessions (**Fig. 3C**), participants decide whether to join a 'ferry rating club', or participants who opted out did not rate the ferry and did not see a rating dashboard. Within a game (**Fig. 3D**) after each ferry ride, we gave participants a choice whether to contribute ratings or 'free ride' and skip ratings in that turn. Note that this *within game* decision is nested and made by a subset of participants that decided to engage. Overall, this game gave participants more agency by allowing them to repeatedly turn the rules on and off, and to decide to what extent to engage once opted in.

Each participant was assigned to a game where only some of the rules apply. For example, a participant that we assigned to the TFFF group, was exposed to the Leadership rule, and could turn it on by joining the club. Every fifth turn, the game was paused, and participants were presented with ferry club membership choices. For a participant assigned to the TTFF group, both *Leadership rule* and *Payoff rule* were turned on once joining the ferry club. The only exception is with the *Escape rule* -- a privilege that could not be turned off and was designed to discourage both membership and contribution of ferry ratings.

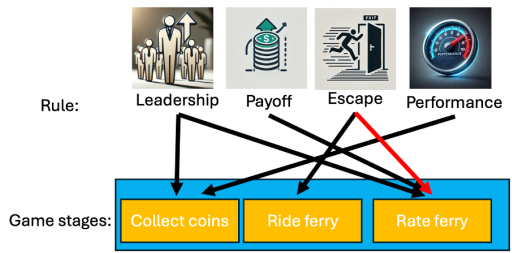
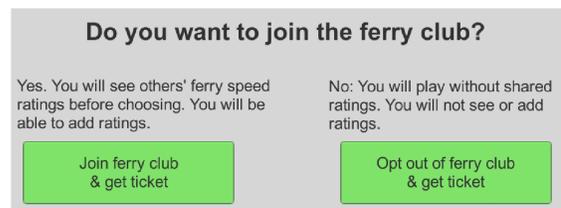
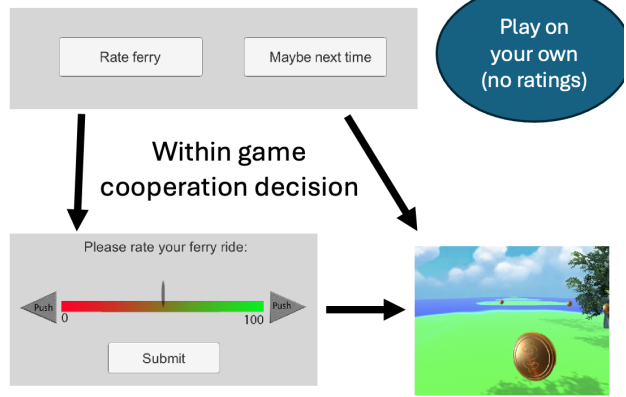

**Figure 3:** *A, Four binary game rules. B, Rules interactions over utilities and game stags. C, User interface for joining membership decision every 5$^{th}$ round (between games). D. User interface for ferry rating cooperation decision, presented to ferry club members immediately after each ferry ride (within game).*

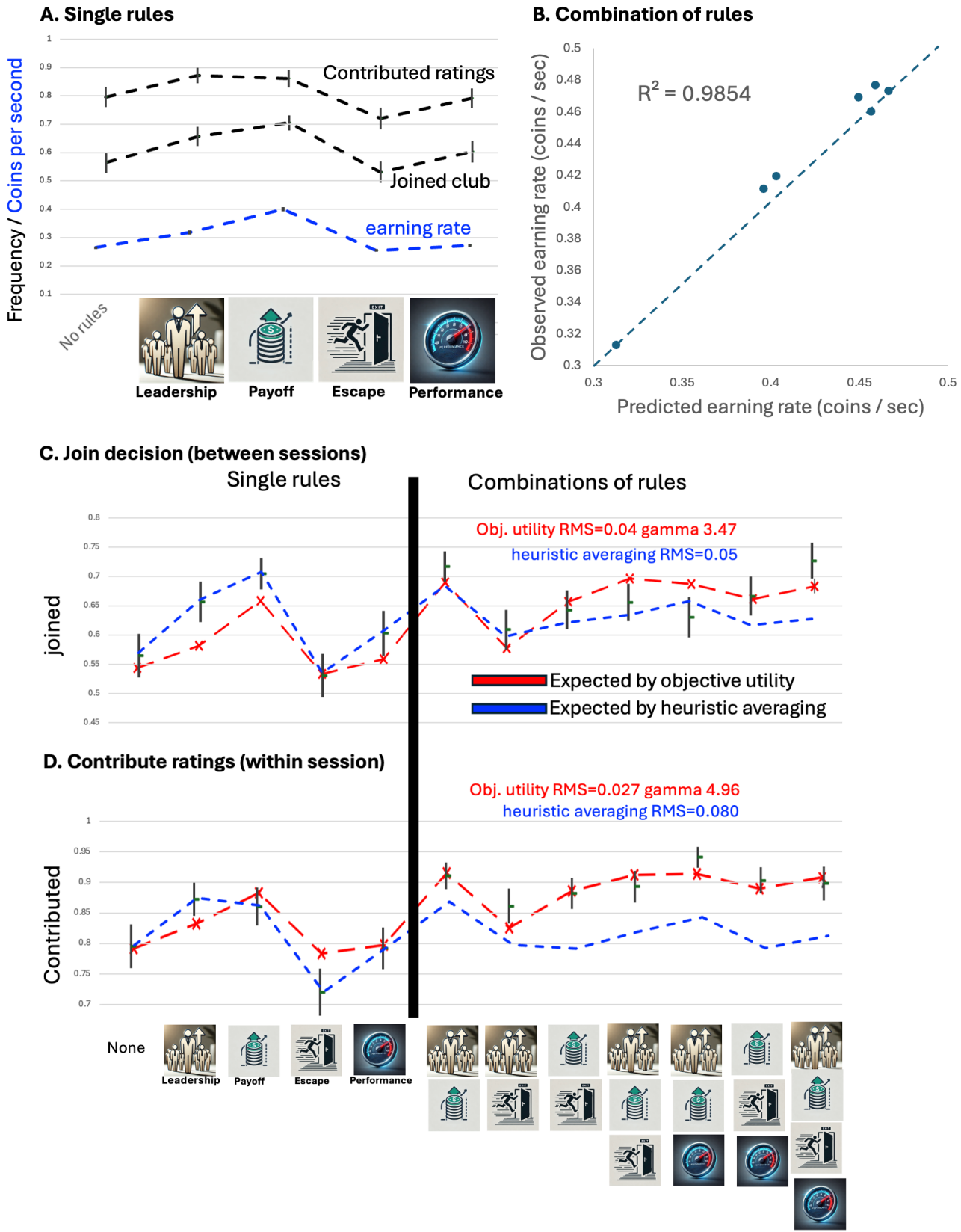

**Figure 4:** *A, Single rules: rates of membership (joined club), contribution of ratings (within members and within the game), and earning rate (coins per seconds). B. Predicted vs. observed cooperation for each combination of rules, given cooperation level with single rules (additive model). C, means and s.e.m. for frequencies of join membership decisions for single rules and combinations of rules. Blue line: predicted by earning rate with single rules. Red: predicted by averaging cooperation level for single rules. D, rates of within game cooperation decisions. Predictions (Red & Blue) as in C.*

We recruited 771 participants, and randomly assigned each participant to a game, with a combination of the four binary engagement rules. **Figure 4A** presents choice frequencies for participants that were assigned to a single rule or no rule. As shown, the frequencies of joining the club, frequencies of contributing ratings, and earning rates (coins per second), show similar trends. For example, the payoff-rule increased earning rate, membership choices (*between games*), and contributing ratings choices (*within game*). As expected, the escape rule reduced membership and reduced the contribution of ratings among members.

**Figure 4B** presents the observed earning rates for combination of rules against the earning prediction based on the earning rates with single rules. As shown, the data are very close to the diagonal confirming that the earnings are indeed additive. We can now test model predictions for the interactions between rules: **Figure 4C** presents the fit to the earning rate (objective utility, model equation (1)), against the prediction of heuristic averaging (model equation (3)). As shown, both objective utility and heuristic averaging models fit the data fairly well, with predictions that are within the error margins for most rules and rule combinations. We used RMS as a yardstick for estimating the distance between the predicted and observed frequencies. As shown, for both models RMSs are small (0.04 and 0.05).

Interestingly, for the *within game* choices (**Fig. 4D**), the objective utility model gave remarkably accurate prediction of choice frequencies, with model estimates falling within the error margin of 10 out of the 12 treatment groups, with gamma of 4.96 indicating high influence, and very low RMS distance of 0.027 from the observed frequencies. In sharp contrast, the heuristic averaging model predictions, which are by design a perfect fit to the single rules, falls outside the error margins of all the combinations of rules, well below expectation (binomial test p= 0.008).

## Discussion

We first found that when choosing between worlds, participants' decisions could not be predicted by the earning rate. Even subjective utility predictions were no better (and often inferior) to those of naive averaging heuristics. In the second game, however, earning rate predicted membership choices (*between games*) well, but so did the averaging heuristic. Interestingly, for the subset of participants who joined the club, objective utility gave accurate predictions for the frequencies of instantaneous cooperation decisions *within game*. Whereas heuristic averaging (folks' theory) failed, the objective utility model gave an excellent fit to the choice frequencies for both single rules and for combinations of rules.

We did not design the two games to be compatible. In particular, the ferry tax aversion might have derailed the objective utility predictions in the first game, while self-selection may be inflating predictions in the second. But even so, between-game choices in both games, at best, did not show any advantage of objective utility predictions against those of heuristic averaging. Are between-games decisions recruiting different faculties than decisions within them? What is it that made the within-game decisions more consistent with rationality? The subset of engaged participants who joined the club might have played more rationally compared to the general population, either because they were engaged, or because they self-selected the rules(Krebs et al., 2024). An alternative scenario is that the decisions *between games* are less rational because they slow and removed from the game dynamics, whereas *within game* decisions are more rational because they are implicit, faster, and likely to be primed by the recency of game events (Ludwig et al., 2020). Either way, this outcome has interesting implications: modern communication technology enables real-time influence of citizens influence over the functioning of the system via online feedback. The emerging ubiquity of feedback systems show that people are now constantly sharing information about rules of environments that are more likely to foster desired social outcomes.

In our experiments we provide people opportunities to select between worlds that vary in one or several rules, and then to experience the worlds they selected. With this setup we can evaluate how successful people are at forecasting their own collective's behavior under increasingly complex conditions. In choices between-worlds we find evidence that people deviate from rational benchmarks in favor of a simple heuristic. Importantly, however, with instantaneous feedback from engaged participants, cooperation becomes well aligned with the objective utility of the rules. Namely, regardless of the accuracy of feedback ratings (Tchernichovski et al., 2023), one can potentially obtain reliable information about objective utility, simply by looking at the proportion of opt-in (providing ratings) decisions from people who are interacting with utilities and providing feedback about their experience in real time.

Overall, we hope that our study can contribute to the understanding of a transitivity formalism for forced binary decisions between social systems, comparing decisions made within- and between- artificial social worlds. Integrating recent advances in experiment design and institutional theory, should enable new inquiry into multi-level decision-making in complex social environments, towards a cognitive science of self-organized self-governance.